\documentclass[aps,prl,twocolumn,showpacs,letterpaper, superscriptaddress]{revtex4}
\usepackage{graphicx}
\begin{document}

\title{Pressure-Tuned Spin and Charge Ordering in an Itinerant Antiferromagnet}
\author{Yejun Feng}
\author{R. Jaramillo} 
\affiliation{The James Franck Institute and Department of Physics, The University of Chicago, Chicago, IL 60637}
\author{G. Srajer}
\author{J.C. Lang}
\author{Z. Islam}
\affiliation{The Advanced Photon Source, Argonne National Laboratory, Argonne, IL 60439}
\author{M.S. Somayazulu}
\affiliation{Geophysical Laboratory, Carnegie Institute of Washington, Washington, DC 20015}
\author{O.G. Shpyrko}
\affiliation{Center for Nanoscale Materials, Argonne National Laboratory, Argonne, IL 60439}
\author{J. J. Pluth}
\affiliation{CARS and Department of Geophysical Sciences, The University of Chicago, Chicago, IL 60637}
\author{H.-k. Mao}
\affiliation{Geophysical Laboratory, Carnegie Institute of Washington, Washington, DC 20015}
\author{E.D. Isaacs}
\affiliation{The James Franck Institute and Department of Physics, The University of Chicago, Chicago, IL 60637}
\affiliation{Center for Nanoscale Materials, Argonne National Laboratory, Argonne, IL 60439}
\author{G. Aeppli}
\affiliation{London Centre for Nanotechnology and Department of Physics and Astronomy, UCL, London, WC1E 6BT, UK}
\author{T.F. Rosenbaum} \email [Corresponding author.  Email: ]  {tfr@uchicago.edu}
\affiliation{The James Franck Institute and Department of Physics, The University of Chicago, Chicago, IL 60637}
\date{\today}

\begin{abstract}
Elemental chromium orders antiferromagnetically near room temperature, but the ordering temperature can be driven to zero by applying large pressures. We combine diamond anvil cell and synchrotron x-ray diffraction techniques to measure directly the spin and charge order in the pure metal at the approach to its quantum critical point. Both spin and charge order are suppressed exponentially with pressure, well beyond the region where disorder cuts off such a simple evolution, and they maintain a harmonic scaling relationship over decades in scattering intensity. By comparing the development of the order parameter with that of the magnetic wavevector, it is possible to ascribe the destruction of antiferromagnetism to the growth in electron kinetic energy relative to the underlying magnetic exchange interaction.
\end{abstract}



\maketitle

Electrons carry not only charge but also spin, and how magnetic order develops in metals where charge carriers remain itinerant continues to be a central problem in both condensed matter and device physics.  As technology progresses and device dimensions shrink, quantum effects become more pronounced and a variety of potential ground states can emerge with coupled charge, spin and orbital order \cite{dagotto}. These effects are most acute near quantum phase transitions, where magnetism first emerges at the absolute zero of temperature \cite{millis, sachdev}.  In particular, antiferromagnetic coupling between interacting mobile electrons is believed to underlie some of the most profound puzzles in modern metal physics, most notably exotic superconductivity, heavy fermions, and other non-Fermi-liquid phenomena \cite{mathur, custers}. 

Nevertheless, definitive characterization of quantum critical behavior in itinerant magnets has proved elusive. Direct order parameter studies of stoichiometric, itinerant ferromagnets suggest that the quantum phase transition is always first order, shrouding the critical behavior \cite{pfleiderer}.  Quantum critical behavior in itinerant antiferromagnets has been observed using indirect probes such as electrical transport and specific heat \cite{custers}, but no direct studies of the order parameter of stoichiometric antiferromagnets exist.  As a further complication, the effects of chemical doping and substitution are amplified at a quantum phase transition, where materials become ``hypersensitive" to disorder \cite{coleman}. 

Directly observing the emergence of antiferromagnetism in a model stoichiometric system without the application of a symmetry-breaking field or dopant disorder would reveal fundamental aspects about the magnetic order itself. To this end, we present a direct x-ray diffraction study of the spin and charge order parameters in elemental chromium, the archetypical itinerant spin-density-wave (SDW) antiferromagnet, as the magnetic order is suppressed with pressure towards its quantum phase transition. Cr is attractive as a model system \cite{overhauser_fedders, rice, young, yeh, lee} and is particularly amenable to theoretical exposition given its simple bcc crystal lattice and well-understood Fermi surface, disorder-free magnetic tunability \cite{mcwhan}, and availability in high purity, single crystal form \cite{method}. 

\begin{figure} 
\begin{center}
\includegraphics[width=3.375in]{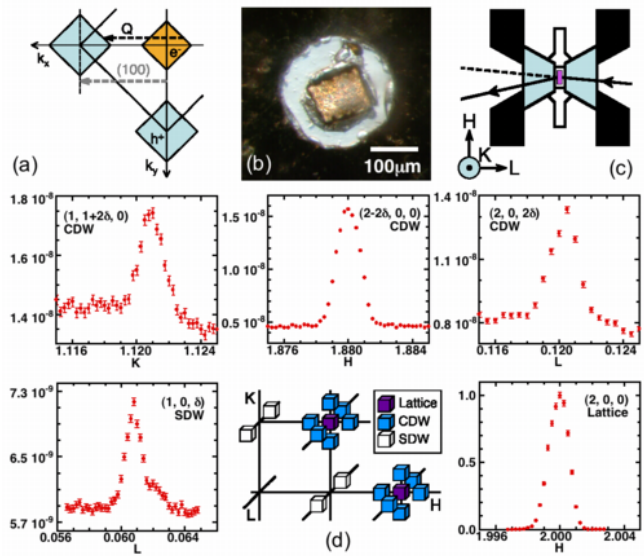}
\caption{(color online). X-ray diffraction measurement of incommensurate SDW and CDW in a diamond anvil cell. (a) Schematic of Cr's Brillouin zone; only the two \textit{d}-like bands involved with magnetic ordering are shown for clarity. The nesting vector $\mathrm{Q} = 1-\delta$ is incommensurate with the lattice, and depends on the relative size of the electron (orange) and hole (light blue) octahedra. (b) Typical view through the pressure chamber shows a single-crystal Cr sample, roughly $120\times100\times40 \mu m^3$ in size.  (c) Schematic of diffraction in the transmission geometry through the cell. Cr crystal (purple) is prepared with one cubic direction (\textit{e.g.} L) aligned parallel to the cell axis. (d) High-resolution and high-sensitivity monochromatic x-ray diffraction from Cr at 8 K and 4.0 GPa. Schematic (not to scale) shows diffraction peaks that are accessible given our cell geometry, all of which are within or very close to the H-K plane. SDW and CDW Bragg reflections appear as satellites around cubic reciprocal lattice points \cite{hill}. Five panels show typical raw scans of the lattice, CDW, and SDW Bragg peaks. All data have been normalized to the (2,0,0) peak intensity.}
\end{center}
\end{figure}

The SDW in Cr is stabilized by two nested sheets of Fermi surface, which are eliminated in the magnetic phase by the formation of an exchange-split energy gap \cite{overhauser_fedders}. The SDW is modulated by a wavevector $\mathrm{Q}$ which is selected by the nesting condition and is slightly incommensurate with the crystal lattice (Fig.~1(a)). $\mathrm{Q}$ may lie with equal probability along any of the three cubic axes, defining three orthogonal Q-domains. Below the N\'eel temperature, $\textit{T}_{N}$= 311 K, and above the spin-flip temperature, $\textit{T}_{SF}$ = 123 K, the SDW is transverse and the spins preferentially lie along either cubic axis perpendicular to Q, defining two possible S-domains; below  $\textit{T}_{SF}$  the SDW is longitudinal \cite{hill}. The SDW is accompanied by a charge density wave (CDW), which is modulated by $\mathrm{2Q}$ and is usually thought of as the second harmonic of the SDW \cite{young}. This harmonic relationship between spin and charge is consistent with the $\textit{I}_{CDW} \propto \textit{I}_{SDW}^2$ scaling (where $\textit{I}$ is scattering intensity) that is observed as a function of \textit{T} \cite{hill}, but has not been tested as the ground state is tuned towards the quantum critical point.

Early transport studies of the pressure $\textit{P}$ dependence of antiferromagnetism in Cr placed the $\textit{T}$ = 0 phase transition above 8 GPa \cite{mcwhan}, necessitating the use of a diamond anvil cell.  The capability of probing the spin (SDW) and charge (CDW) order parameters using x-ray diffraction has been demonstrated previously at ambient $\textit{P}$ \cite{hill, strempfer}. Here we extend such measurements to high $\textit{P}$ at liquid helium temperatures (Fig.~1(b,~c), \cite{method}). Our sensitivity is approximately $5\times10^{-10}$ relative to the BCC Bragg intensity ($1/10^{th}$ of the background), which is sufficient for following the spin and charge order parameters into the quantum critical region.

Precise measurement of the order parameters is complicated by the presence of Q- and S-domains; all of which must be accounted for if the SDW and CDW diffraction intensities are to be properly normalized. By measuring CDW peaks corresponding to all three Q-domain types (Fig.~1(d) schematic) and taking into account the atomic form factor \cite{diana} and the appropriate strain wave cross section \cite{hill}, we are able to calculate the Q-domain distribution, presented in Fig.~2(a). The unpredictable evolution of the domain distribution with pressure underscores the need to account for the entire domain structure when measuring the order parameters. Furthermore, the presence of Q-domains along all three cubic axes attests to the quasi-hydrostatic nature of the pressure environment. A uniaxial compressive stress along the DAC axis would favor domains with Q along the L-direction, owing to the orthorhombic symmetry of the crystal in the SDW phase. Even for the pressure with the most imbalanced domain distribution (4.1 GPa), the uniaxial stress is estimated to be less than 0.02 GPa based on the strain necessary to force a single Q-domain state \cite{steinitz}.

We find that the longitudinal phase is completely suppressed above $P \approx 1\> \mathrm{GPa}$ at $T = 8\> \mathrm{K}$, so that all high-pressure measurements presented here are made in the transverse phase. Therefore, it is necessary to measure two inequivalent SDW reflections (such as (1,0,$\pm\delta$) and (0,1,$\pm\delta$)) in order to determine the S-domain distribution. The SDW ordered moment is then calculated from the equation:
\begin{equation}
I_{SDW}/I_{Lattice}=(\hbar\omega/m_{e}c^2)^2(f_{m}/f)^2(\mu/N)^2,
\end{equation}
where $\hbar\omega$ is the x-ray energy, $\textit{f}_{m}$ and \textit{f} are the magnetic \cite{strempfer} and atomic \cite{diana} form factors, \textit{N} is the number of electrons per site, $\mu$ is the (r.m.s.) ordered moment per atom in units of $\mu_{B}$ \cite{hill}, and $\textit{I}_{SDW}$ and $\textit{I}_{Lattice}$ are the (properly normalized) SDW and lattice diffraction intensities. Our measured ordered moment at ambient \textit{P} is $0.39 \pm 0.02 \mu_B$, consistent with the accepted value of $\mu_0=0.41$ \cite{hill}.

We plot in Fig. 2(b) the evolution of the SDW and CDW diffraction intensities with \textit{P} at \textit{T} = 8 K. Both order parameters are seen to scale exponentially with pressure over the entire measurement range. The quadratic scaling between $\textit{I}_{CDW}$ and $\textit{I}_{SDW}$ holds as a function of pressure, indicating that the coupling of the spins to the charge is not altered by varying the lattice constant. The exponential dependence of the ordered magnetic moment on pressure can be understood within the framework of the two band model of a nested SDW, which for the case of perfect nesting is analogous to a BCS superconductor \cite{overhauser_fedders}. To the extent that the exchange potential between the two nested sheets of Fermi surface is a constant, the model predicts that $\mu \propto g_0$, where $2g_0$ is the energy gap responsible for eliminating the magnetic Fermi surfaces \cite{overhauser_fedders}. We therefore apply the expression for the gap \cite{overhauser_fedders}:
\begin{equation}
g_0\propto \exp(-2\pi^2v/\gamma^2 \overline V k_c^2 ) \equiv \exp(-1/\lambda),
\end{equation}
to the experimentally determined ordered moment. Here $\gamma$ is an average exchange integral, $\overline V$ is an average Coulomb potential, $4\pi k_c^2$ is the Fermi surface area of the magnetic bands, and $v$ is an average Fermi velocity. Comparing our data to Eq.~2 we see that $1/\lambda$, the ratio of kinetic energy density $t = v/k_c^2$ to exchange potential density $J = \gamma^2 \overline V$ in reciprocal space, varies linearly with \textit{P}. The exponential decrease in ordered moment with pressure is consistent with the similar decrease in $T_N \propto \exp(C_{T} \cdot\Delta a/a_0)$ with $C_{T} = 93$ \cite{mcwhan}. We conclude that the suppression of antiferromagnetic order with pressure at the approach to the quantum critical point is described simply by a BCS-like linear relationship between $T_N$ and $g_0$.

\begin{figure}
\begin{center}
\includegraphics[width=3.375in]{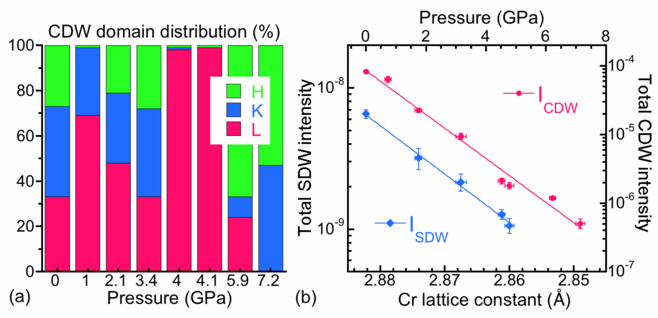}
\caption{(color online).  (a) CDW Q-domain distributions for samples at different pressures and 8 K; ambient \textit{P} data was taken at 130 K in the transverse phase to be consistent with the high pressure points.  (b) SDW and CDW diffraction intensities $I_{SDW} = I_{(1,0,\delta)} / I_{200}$ and $I_{CDW} = I _{(2Q,0,0)} / I_{200}$ as a function of pressure at \textit{T} = 8 K. Data have been normalized to the (2,0,0) lattice Bragg peak intensity and account for Q- and S-domain distributions. Both intensities are suppressed exponentially with the BCS form $I_X \propto \exp(C_Xá\Delta a/a_0)$, where $a_0=2.8820$ \AA$ $ is the lattice constant at ambient \textit{P} and 8 K; $C_{SDW} = 227 \pm 10$, $C_{CDW} = 457 \pm 25$. The $I_{CDW} \propto I_{SDW}^2$ scaling between the CDW and SDW intensities at \textit{T} = 8 K is consistent with scaling seen at ambient pressure where \textit{T} is varied \cite{hill}.}
\end{center}
\end{figure}

\begin{figure}
\begin{center}
\includegraphics[width=3.0in]{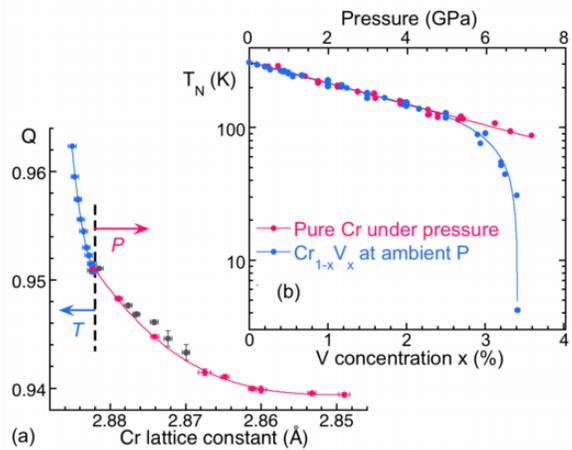}
\caption{(color online). (a) Evolution of the SDW wavevector $\mathrm{Q}$ (in units of $2\pi/\textit{a}(\textit{P}))$ with temperature and pressure.  $\mathrm{Q}$ becomes independent of \textit{P} at the approach to the quantum phase transition. \textit{T} dependence at ambient \textit{P} (blue) and \textit{P} dependence at 92 K (gray) were measured with energy dispersive x-ray diffraction \cite{feng}; \textit{P} dependence at 8 K (red) was measured with a monochromatic technique. Solid lines are guides to the eye. (b) Pressure (red) and Vanadium doping (blue) track until disorder disrupts the BCS behavior. $T_N$ for pure Cr under pressure is derived by using a linear relationship $T_N(P)= \mu(P)/\mu_0 \cdot 311$ K, and from Ref. \cite{mcwhan} with a pressure scale reduced by a factor of 1.24 to account for the difference in pressure calibrations. $T_N$ for Cr$_{1-x}$V$_{x}$ are collected from Ref. \cite{yeh, lee, koehler, CrV} and references 15-17 within \cite{lee} for dopant concentrations up to $x_c = 3.42\%$. The conversion between \textit{P} and \textit{x} is set by the collapse of all the doping data for $x \leq 2.5 \%$ onto the same exponential curve of pure Cr data under pressure, giving $dP/dx = 1.99$ GPa/$\%$.}
\end{center}
\end{figure}

Pressure can suppress the gap $\textit{g}_{0}$ principally through either the magnetic exchange coupling or the Fermi surface geometry of the magnetic bands. We separate these effects by studying the \textit{P} dependence of the SDW wavevector $\mathrm{Q}$, which reflects the evolution of the bandstructure and the nesting condition. We compare in Fig.~3(a) the \textit{P} dependence of $\mathrm{Q}$ at low \textit{T} and its \textit{T} dependence at ambient \textit{P}. Q(\textit{P}) appears to flatten off with increased pressure, approaching a constant for $\textit{P}> 4\> \mathrm{GPa}$. Given that the observed exponential dependence of $\mu$ on \textit{P} continues to at least 7.2 GPa, this demonstrates that the change in $\mathrm{Q}$ is not directly related to the evolution of $g_0\propto \exp(-1/\lambda)$. Changes in $\mathrm{Q}$ at \textit{T} = 0 reflect changes in the geometry of the underlying paramagnetic nested Fermi surfaces, resulting from a redistribution of electrons in reciprocal space; for a rigid bandstructure $\mathrm{Q}$ would remain constant with pressure. The data show that the geometry of the nesting bands is essentially pressure independent, approaching the rigid limit for $\Delta a/a_{0} > 1\%$. 

It is important to consider the possibility that imperfect nesting may alter the physics. In such a case, the the spin response function at $\chi_{0}(q=Q)$ would exhibit a broad maximum rather than a singular cusp \cite{Windsor}, and $g$ would be renormalized away from $g_{0}$ \cite{rice}. However, this posited renormalization is inconsistent with the observed exponential decrease in $g$ with pressure, indicating that the giga-Pascal pressures applied in our experiment do not warp the bandstructure to the extent of significantly affecting the nesting condition. 

The contrast between the physics of clean and disordered materials is brought into sharp relief in Fig.~3(b). It is possible to suppress $\textit{T}_{N}$ not only with pressure, but also by doping with V \cite{yeh, CrV}. The two techniques track over a wide range, but chemical doping, which introduces both disorder and change in the average electron count, drives the system away from a BCS form for $\textit{T}_{N} \leq 2/5\textit{T}_{N}(\textit{x}=0,\textit{P}=0)$. The simple description of the evolution of magnetism in terms of the BCS energy gap is cut off by disorder, hastening the onset of the proximate quantum phase transition. Furthermore, for 2.5$\%$ V doping ($\textit{T}_{N} \approx 125$ K) the low-\textit{T} value of $\mathrm{Q}$ moves to 0.922 and $\mu$ is suppressed to 0.24$\mu_{0}$ \cite{koehler}. By comparison, compressing pure Cr drives $\mu$ to 0.24$\mu_{0}$ when $\mathrm{Q}$ = 0.9394, only 1/3 the change in $\mathrm{Q}$.  The evolution of $\mathrm{Q}$(\textit{x}) for Cr$_{1-x}$V$_{x}$ \cite{koehler, fawcett} does not level off for larger values of \textit{x} as does $\mathrm{Q}$(\textit{P}). Therefore, we find that chemical doping in the form of deviations from stoichiometry alters Cr's band structure to considerably larger degree than applying pressure. At a minimum this makes separating competing physical effects more difficult and at a maximum may point to the mechanism by which deviations from pure BCS-like behavior ensue.

The direct measurement of the SDW and CDW order parameters at temperatures where thermal fluctuations are frozen out, combined with the flattening off of the SDW wavevector at the approach to the quantum critical point, reveal that it is the ratio of magnetic exchange to kinetic energy, $\lambda = J/t$, which drives the quantum phase transition in pure Cr. A quantitative estimate for $\lambda$ can be obtained from recalling that for three-dimensional Fermi liquids we expect the total kinetic energy of the electrons to scale like $a^{-5}$, which increases strongly under pressure as a quantum confinement effect.  From the data up to \textit{P} = 7.2 GPa, the fractional change in the kinetic energy $\Delta t/t_0 = -5\Delta a/a_0 = 0.06$ which, together with the measured $\Delta(1/\lambda) = 1.3$, leads to $\lambda = J/t \sim 0.05$. We have assumed that the exchange interactions are not sensitive to the lattice constant, and indeed, if anything, they should become stronger with pressure, which would imply an even larger kinetic energy contribution to the change in $\lambda$. In this picture, the magnetic order is destabilized by the increase in the kinetic energy due to quantum confinement.

Even with $\lambda$ small, the spin interactions can be robust. Inelastic neutron scattering studies demonstrate that spin-spin correlations survive up to surprisingly high energies and temperatures \cite{hayden}, and the Hall coefficient retains a strong temperature dependence well above the N\'eel transition \cite{yeh}. The intrinsic exchange-driven pairing potential is thus strong in Cr, but its ground state evolution still can be modeled by the BCS-like gap solution given the high density of carriers and consequent high degree of pair overlap \cite{rink}. Of course, the BCS form for the order parameter only approaches zero asymptotically. As with chemical doping (Fig.~3(b)), the curve is likely to turn over and assume a critical form at sufficiently large \textit{P}. Alternate broken symmetries such as superconductivity become a possibility if disorder is sufficiently weak, as does the survival of a purely second-order quantum phase transition, depending on the strength of the coupling of the harmonics to the primary SDW \cite{young}.

In summary, we have performed a direct study of spin and charge ordering in the stoichiometric itinerant antiferromagnet Cr as the magnetic order is suppressed with pressure. We find that this suppression is an effect of quantum confinement, the ground state evolves in accordance with the BCS weak-coupling theory, and the spin and charge degrees of freedom have a persistent harmonic relationship. The low-\textit{T} behavior of the pure element differs dramatically from that of the family of doped systems approaching the quantum phase transition.

The work at the University of Chicago was under NSF Grant No. DMR-0534296.   G.A. acknowledges a Royal Society Wolfson Research Merit Award. R.J. thanks a NSF Graduate Research Fellowship. Use of APS is supported by the U.S. DOE-BES, under Contract No. NE-AC02-06CH11357. Use of HPCAT is supported by DOE-BES, CDAC, NSF, and the W.M. Keck Foundation.

\end{document}